# COMPUTER SCIENCE PROGRAMS, GOALS, STUDENT LEARNING OUTCOMES AND THEIR ASSESSMENT

**Vijay Kanabar, PhD, Anatoly Temkin, PhD**

**Boston University, Metropolitan College, Computer Science**

*Abstract*: In this paper we describe the process used by MET Computer Science to identify programs, goals, and student learning outcomes for all our programs and graduate certificates. We illustrate how we started the process of assessing learning outcomes for our programs and present actual assessment results from the data collected for two programs that went through accreditation--it describes sample direct and indirect assessment data from two of our programs.

*ACM Classification Keywords*: K.3.2 Computer and Information Science Education, Curriculum, Information Systems education, self-assessment, K.3.m Accreditation

## Introduction

Established in 1979, the Department of Computer Science at Metropolitan College (MET) is the longest-running computer science department at Boston University. MET's Computer Science department remains a leader in a number of state-of-the-art IT areas, such as information security, computer networks, computer information systems, financial informatics, digital forensics, and health informatics. A regional and national leader in information security education for almost a decade, the curriculum is certified by the Committee on National Security Systems (CNSS).

World-class IT programs are driven by exceptional faculty whose scholarly accomplishments and unique industry experience place them at the top of their field. Adhering to the high academic standards of BU, these dedicated scholars are involved in research projects in areas such as novel Internet architectures, smartphone applications, information assurance, and biomedical informatics. Most importantly, professors are fully engaged with their students, following their progress, maintaining awareness of what's going on in their lives and careers, and providing the support they need. An academic advisor is also available to help students make the best decisions about classes.

Part-time and Online study options offer convenient times, locations, and delivery methods, ensuring that dedicated professionals have the opportunity to experience a rigorous academic environment while pursuing full-time careers. At the same time, we remain a popular choice for a number of international students interested in the BU experience, and in continuing their academic development in computer science and information technology.

Late in 2013 we were informed about a university wide initiative to align all educational programs with the expectations of New England Association of Schools & Colleges, Inc. (NEASC), the nation's oldest regional accrediting association whose mission is the establishment and maintenance of high standards for all levels of education accreditation bodies. While this effort was a huge challenge, we undertook it with a great sense of responsibility because upon graduation our students must have the ability to analyze problems efficiently and possess the tools of the prestigious degree credential from Boston University. The doors to further higher education in academia and careers in the industry open up for our graduates if we do a good job.

## Program Learning Outcomes Assessment Structure

Before we introduce the process of designing program learning outcomes, we must answer the question of why the process is important. According the university, the "Program Learning Outcomes Assessment" provides faculty a means to ask a fundamental question about the programs they design and teach: by completing a given set of courses and other requirements, do students actually acquire the particular knowledge, skills, habits of mind, and attitudes faculty intend? If not—or if not fully enough—what pedagogical and curricular reforms can be undertaken to improve student learning? A similar question can be asked of the co-curricular and extra-curricular programs that contribute so significantly to a well-rounded education at BU". (BU, 2015)

Program Learning Outcomes Assessment is overseen by the Council of Deans and coordinated and facilitated by the Associate Provost for Undergraduate Affairs and the Associate Provost for Graduate Affairs. University Working Committees, comprised of representatives from the Schools and Colleges, meet regularly with the Associate Provosts to share information and best practices, and to coordinate the University's effort. The department chairs and directors work closely with the representatives to move the process forward at the departmental level.

Regular assessment tutorials and workshops are provided to help faculty coordinators understand the process as well as seek answers for questions from the experts.

## Learning Outcomes Process

Before we introduce the process of designing program learning outcomes, we would like to introduce the process from a higher ground. An outcomes-based approach to education clearly specifies what students are expected to learn and arranges the curriculum such that these intended outcomes are achieved (Harden, R.M. 2007a). Learning outcomes mapping and an effectively aligned and integrated curriculum, in which instructional activities and assessment strategies are explicitly linked to course-specific and degree-level learning outcomes, is at the heart of the learning outcomes process; such mapping is usually always tied to institutional and state-defined degree level expectations (Kenny, N & Desmarais, S, 2012). Figure 1 illustrates how outcomes based alignment flows from the classroom and course to the State/Province (which is very important in the case of a state funded university/college).

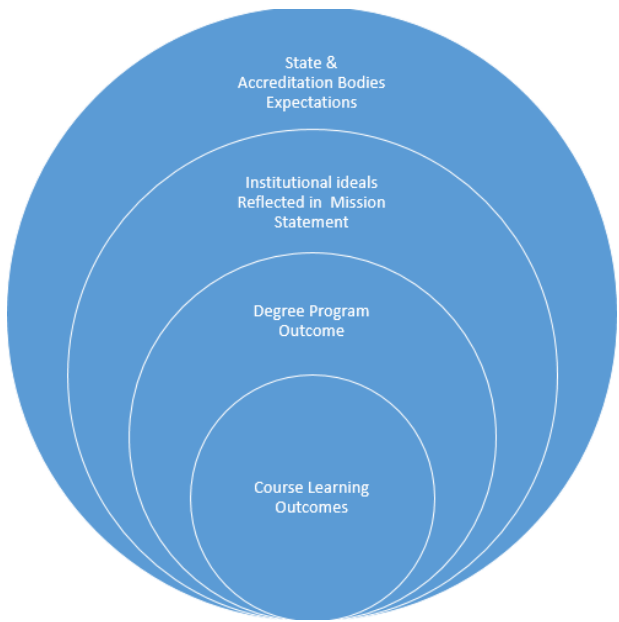

Fig. 1: Graphical representation of outcomes-based curriculum alignment

*Computer Science Programs, Goals, Student Learning Outcomes and their Assessment*

Let us consider the need for alignment or compatibility as reflected in the Figure 1. Computer Science curricula are embedded within colleges; for example, our Computer Science program curriculum is embedded in Metropolitan College. Since MET has specific program learning goals they must be accommodated by the curriculum as well. In our case the MET college goal is to provide education to adult learners; in most cases, they are working adults who don't have time to take day classes. Such students are experienced and interested in current state of the art practitioner knowledge. So our program objectives of the CS curriculum is aligned with the mission statement of the college.

In designing educational objectives we followed the following distinct processes. Figure 2 illustrates the key steps (Task Force on PM Curricula, 2015):

Step 1: Identify Program Goals for each Degree: A goal is a statement of general outcome. It is a broad definition of student competence. It answers the question, "What will the students learn from a program?" It defines educational expectations of a program. In the examples below we have stated the program goals in terms of what we want our students to be able to do upon graduation.

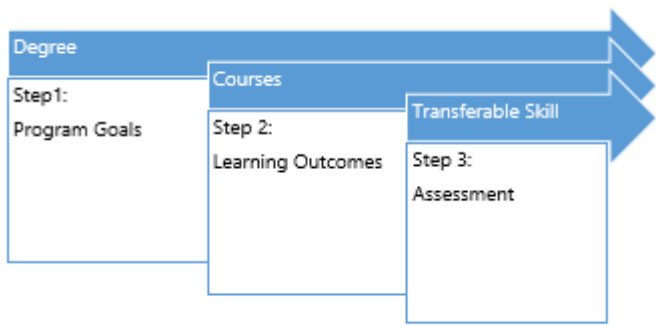

Figure 2: Learning Outcomes Process adopted

Step 2: Identifying Course Objectives and Learning Outcomes: An objective is a statement summarizing specific course content. It answers the question, "What will the students learn from a course?" Learning objectives should map to program goals. Furthermore, a detailed description of what the student must be able to do at the completion of a course is also addressed within this context. When writing outcomes it is useful to identify transferable skills.

Step 3: Assessment. Here we answer the question, "Did they learn?" This involves direct and indirect assessment of the learning outcomes.

## Key Steps

The key steps followed were prescribed by the provost's office. These mapped with the process indicated in Figure 2 well but are action oriented. The first three steps are shown in Figure 3 and analyzed next.

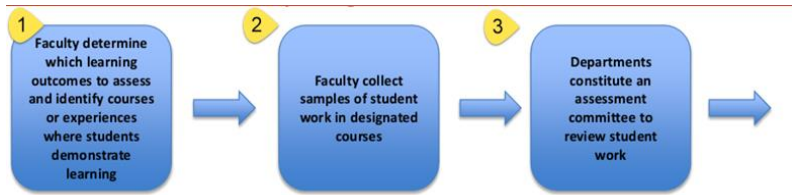

Figure 3: Learning Outcomes Key Steps Followed

***Step 1 – Faculty Determine Learning Outcomes:*** As indicated earlier, we studied the college goal as well as the mission of the university

*Boston University Mission*

"We remain dedicated to our founding principles: that higher education should be accessible to all and that research, scholarship, artistic creation, and professional practice should be conducted in the service of the wider community—local and international. These principles endure in the

*Computer Science Programs, Goals, Student Learning Outcomes and their Assessment*

University's insistence on the value of diversity, in its tradition and standards of excellence, and in its dynamic engagement with the City of Boston and the world."

*Metropolitan College Mission*

Our mission, since 1965, has been to ensure that the resources of a leading research university are accessible to the community and the world, while providing students with high-quality, academic degree and certificate programs, first-rate faculty, and flexible modes of study.

Within this context we identified the program goals for all our degrees.

See the program goal for the security concentration below.

1. Demonstrate advanced knowledge of information security concepts, governance, biometric systems, database systems security, as well as network security and cryptography.
2. Demonstrate proficiency in risk management, such as asset assessments, architectural solutions, modeling, and design.
3. Demonstrate competence in security policies, processes, technology and operations.

Similarly three primary outcomes were identified for all degrees, concentrations and graduate certificates. See example in the next section.

### *Step 2: Identifying the Data for Assessment*

Our next step was to identify data for assessment. This required us to classify direct and indirect assessment data. A good summary of the concepts is provided next.

**"Direct Assessment:** This involves looking at student performance by examining samples of student work. This assessment may examine student outcomes from a given course, from a degree program, or from the University overall (as in achieving University General Education

Goals). Examples of the work to be assessed are: targeted objectives exhibited on final exams questions, student papers or presentations assessed for achievement of course or program goal, student portfolios assessed for achievement of course, program, or University goals, or licensure exams for professional programs." (Skidmore, 2015)

We illustrate the instruments we adopted below.

| | Instrument | Description |
|---|---|---|
| Direct → | Capstone Projects | Projects or Assignments in Capstone Courses |
| Direct → | Exams & Assignments | Tests |

Note that two of the programs, Computer Science and Project Management, have gone through accreditation. This is regarded as "Direct Assessment". For instance, the GAC accreditation completed recently involved a team of external four members who did an outstanding job reviewing all deliverables in the program for assessment purposes.

**Indirect Assessment:** Skidmore defines indirect assessment as follows: "… gathering information about student learning by looking at indicators of learning other than student work output. This assessment approach is intended to find out about the quality of the learning process by getting feedback from the student or other persons who may provide relevant information. It may use surveys of employers, exit interviews of graduates, focus groups..." (Skidmore, 2015). We illustrate the instruments we adopted below.

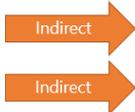

We have an internship program and the internship supervisor at the company is asked to comment on the student's learning upon completion of the internship. This is also an example of indirect assessment. An Alumni survey is done annually and this provides substantial indirect assessment data.

*Which is better? Direct or indirect?*

As a note of comparison we quote, "…both of these assessment approaches provide useful information in improving student learning. Indirect assessment can gives us immediate feedback which can be employed in a course to bring direct improvement to student learning. Unfortunately indirect assessment does not provide reliable evidence that learning objectives have been achieved. The use of surveys and focus groups may lead to improvements in a program but do not directly provide evidence of student learning." (Skidmore, 2015)

### Step 3: Department Constitutes Assessments Committee.

Under the directions of the chair, an assessments committee was formed that consisted of 3 to 4 faculty members. The chair of the department and the curriculum assessment director were present in all the committees. Furthermore, one key member was identified as a curriculum coordinator for each concentration. Example: MS Security – 4 faculty members were involved. The first task of the assessments committee was to discuss the assessments process and timeline.

## The Assessment Process

The faculty were asked to identify deliverables for direct assessment in accordance to university recommendation and best practices widely acknowledged in literature (Gaulden, S., 2010), (Suskie, L. A., 2009), (Maki, P., 2004) & (Walvoord, B., 2010):

- Writing samples from term projects
- Analysis of data pertaining to value added gains to the students' learning experiences by comparing entry and exit surveys
- Locally designed quizzes, tests, with assessments in mind.
- Portfolio artifacts (these artifacts could be designed for introductory, working, or professional portfolios)
- Capstone projects (these could include research papers, presentations, theses, dissertations, oral defenses, exhibitions, or performances)
- Evaluation of case studies
- Team/group projects and presentations
- Passing licensure/certification exams
- Internships reports
- Professional/content-related experiences engaging students in hands-on experiences
- Data about skills in the workplace rated by employers
- Online course discussions analyzed independently from the facilitators evaluations.

We did not focus extensively on indirect measures of student learning in our first assessment, though the Dean's office was willing to provide us data pertaining to graduation of students and the results of alumni surveys. Regardless, we requested all faculty to capture and archive indirect assessment data for consideration by the assessment committee, the following:

- Information about honors, awards, scholarships
- Any form of recognition earned by current students and alumni, employment of graduate students into computing career positions
- Acceptance into a relevant doctoral programs.

Based on the recommendations from a workshop at the Provost's office, we created a template for indirect assessment. This is described in the next section.

## Results from Assessment

In this section we will describe some early results from our assessment of two programs, MS Computer Science with a concentration in Security, and MS Computer Information Systems with a concentration in Information Technology Project Management.

**Direct Assessment Summary:**

The direct assessment report summary for all concentrations is similar to the following:

**Security concentration:**

> Direct assessment of deliverables in key security courses was conducted by the assessments committee in Sept 2015. This approach involved the participation and evaluation of student work by someone other than the instructor. Select deliverables, i.e., assignments and term projects were reviewed by a team. All assessed information and action plan are archived in a departmental network folder.

Extracts from some specific assessments are shown below:

Minutes from a Direct Assessment Deliberation:

The following minutes capture the deliberations that took place within the context of reviewing capstone courses. The participants were Prof. Stuart Jacobs, Prof. Vijay Kanabar, and Prof. Anatoly Temkin. A random sample of assignments and term projects were reviewed.

The following comments pertain to all written deliverables.

- Proper compliance with APA style guide regarding citations and references.
- Style of writing should be succinct.
- Students should present ideas in a coherent and logical manner.
- Ability to professionally articulate the concepts and methods applicable to the issues under discussion.
- Improve opportunities for students to express ideas in a professional manner.
- Students should be able to communicate effectively to non-technical professionals.
- Student should be able to communicate professionally both verbally and in writing.
- Student should have the opportunity to make 30 seconds elevator speeches.
- PowerPoint presentations should be sound and logical. The slides should leave the reader with the conclusions that the presenter wants to make.

Addition technical comments on the Capstone Crypto Project:

- A direct technical assessment of the final project was conducted. It required each student to utilize a crypto analytic algorithm developed throughout the course for recording plaintext from ciphertext.

*Computer Science Programs, Goals, Student Learning Outcomes and their Assessment*

- Although most students succeeded in this project, a few opted for the more challenging ciphertext problems.
- We assessed students in this key security course as having achieved "*High Competency*". The highest degree of possible rating is not "*Exceptional*".

The action plan is an important component of direct assessment. We determined the following action plan for the security concentration.

An action plan was identified to enhance student competency.

- Opportunity to make the final project more challenging. Focus on algorithm performance.
- Have the students opt for the more challenging problems; they would be better prepared for dealing with real world situations.
- We need to provide increased tutorial support for effective technical written and verbal communication.
- We need to provide increased focus on remedial explanation of APA style requirements.
- We need to provide students the opportunity to make effective 30-second elevator speeches
- Modest curriculum changes have been targeted to address the above opportunities. This will be discussed at a Department meeting to get wider input.

**Indirect Assessment of the Security Concentration**

As indicated earlier, we created a survey instrument to assess indirectly the learning outcomes of students graduating from all computer science programs. The survey instrument bluntly asked students to rate how competent they felt about their experience in the program as it pertains to the program learning outcomes. It also had several open-ended questions.

The survey can be found at: http://blogs.bu.edu/metcs/the-survey-questions/assessment-computer-science-security-concentration/

We would like to note that the survey was anonymous and no student names or identifying information was collected.

The students were asked to rate the program goals from 0 to 6

> 0 = None
> 1 = Very low
> 2 = Low
> 3 = Medium
> 4 = High
> 5 = Very High
> 6 = Excellent

The open ended questions yield a lot of useful information.

> Did you acquire other competencies in our program other than those addressed above for example, project management, and communication?

Here the students were asked to Rate: 0 to 6 using the same scale described above and were also asked provide open-ended comments.

Students were also asked to provide some open-ended comments. Some specific comments provided by students included the following:

- We should have more computer lab rooms available
- Useful discussions in class. I appreciate that students are practitioners and have lots of experience to share in the class.
- Very helpful that classes are taught by practicing computer science professors.

*Computer Science Programs, Goals, Student Learning Outcomes and their Assessment*

It should be noted that students also sent us information about employment. Such information is stored in the database. Actual samples anonymized are shown below:

> RK - Job at EMC – Corporate Systems Engineer – Graduated Dec. 2014 and immediately employed.
>
> JL - Offered a Software Engineer (SDE) position in Amazon Seattle HQ – AWS division. Graduated and immediately employed – 2015.

### Pre and Post test

Pre- and post- tests yield useful information of learning taking place in the program. The following is a sample of the pre and post test questions given to the students. Students were asked to rate their knowledge before they joined the program and that confidence on the topic at the end of the program.

Rate your academic background in fundamentals of cryptography.

> Before: ____
>
> After: ____

Rate your understanding of how the RSA and El Gamal ciphers work.

> Before: ____
>
> After: ____

Rate your knowledge of factorization attacks and algorithms to compute discrete logs in cyclic groups.

> Before: ____
>
> After: ____

Rate your understanding of how the key management system works.

      Before: \_\_\_\_

      After: \_\_\_\_

Rate your knowledge of random number generators and algorithms for primality checking.

      Before: \_\_\_\_

      After: \_\_\_\_

We illustrate survey results and analysis for the project management specialization in Figure 4.

### Results from Assessment of the MS CIS IT PM Program

A Student Survey to Measure Changes in Experience, Knowledge, and Competency in Project Management Core Courses was conducted in 2013-2014. Pre-tests and Post-tests analytics were conducted to measure knowledge gained in a specialization core (Kanabar, et al 2014). See sample results in Figure 4.

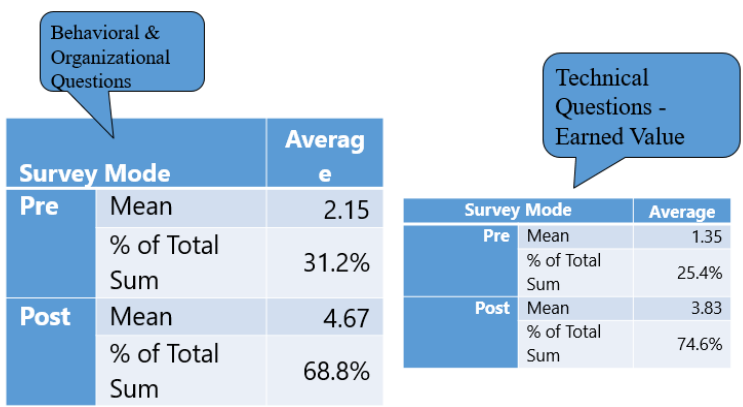

Figure 4: Pre--and Post Test Analysis

*Computer Science Programs, Goals, Student Learning Outcomes and their Assessment*

Once again the assessment process resulted in significant improvements. The degree program was modified -- a new Agile Software Development course was introduced into core. The Software Architecture Course was enhanced and revised. The Distributed Software Development course was strengthened by at least 50%. We hope to assess the program again in 2016 to see if it has resulted in satisfied students.

The provost's office at Boston University provided us a rubric which we used as a benchmark. Our goal at the outset was to "Meet Standards" as shown in the table below. We feel we have exceeded the standards defined in some cases.

**Table 1** Checklists and Standards for Learning Outcomes.

|  | **Meets Standard** |
|---|---|
| Mission | • Clearly states broad aspects of the program's function<br>• Aligned with University Mission |
| Outcomes | • Aligned with and specific to the program's mission<br>• Clearly measurable<br>• Expressed in language that focuses on what students will be able to demonstrate |
| Methods/ Measures | • Content to be assessed fits outcomes<br>• Data collection process is briefly described<br>• Both direct and indirect measures are used |
| Findings | • Findings entered for each measure<br>• Status of finding indicated and clearly described<br>• Appropriate evidence is presented |

| Actions (use of results) | • Action plan is developed from findings and aligned with outcomes<br>• Clearly describes intended improvements<br>• Program shows use of assessment results for improvement |
|---|---|
| Reporting | • Report is complete (all questions answered) and up to date |

In summary, all our programs are aligned with the mission, have clearly defined outcomes, have methods and measures for direct and indirect assessment. This paper describes the actions taken to date and the reporting of our preliminary assessment.

## Conclusion

Program and course assessments have always been conducted in our department. This has worked reasonably well as evidenced by the substantial amount of curriculum enhancements taking place on an annual basis. We have changed courses, introduced new tools or new languages and occasionally entire new degree programs into the curriculum. Subsequently, the quality of delivered education is high and students get jobs successfully. Reviewers of colleges and universities such as US News & World Report rank our programs highly as well. Certainly one reason is that our programs are well designed and executed with due diligence by excellent faculty. In this paper we described our approach to addressing the learning outcomes expectations for our college and specifically the Computer Science department by the university. The process has been very time-consuming, but we believe this effort should result in a sustainable quality program.

*Computer Science Programs, Goals, Student Learning Outcomes and their Assessment*

## Authors' Information

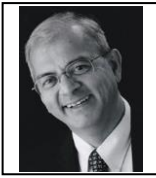
***Vijay KANABAR, PhD, PMP,** Boston University, Metropolitan College, Associate Professor. Email: kanabar@bu.edu*
***Major Fields of Scientific Research:*** *IT Project Management, Curriculum Design and Development, Online education. Web application Development.*

## Authors' Information

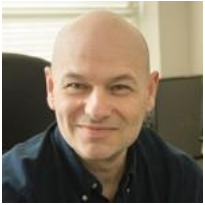
***Dr. Anatoly Temkin,** 808 Commonwealth Ave. Room 250*
*E-mail: temkin@bu.edu*
***Major Fields of Scientific Research:*** *Mathematical Computation, Security, Cryptography*

*Computer Science Programs, Goals, Student Learning Outcomes and their Assessment*